\documentclass[12pt]{article}
\usepackage{amsmath}
\usepackage{graphicx}

\begin{document}
\begin{center}
\section*{Relativistic Calculation of Hadronic Baryon Decays}
\end{center}
\begin{center}
{\bf B. Sengl}, {\bf T. Melde}, and {\bf W. Plessas}\\
\vspace{0.3cm}
Theoretical Physics, Institute for Physics, University of Graz,\\
Universit\"atsplatz 5, A-8010 Graz, Austria
\end{center}
\renewcommand{\abstractname}{}

\begin{abstract}
The description of baryon resonance decays represents a major challenge
of strong interaction physics. We will report on a relativistic 
approach to mesonic decays of light and strange baryon resonances within constituent
quark models. The calculations are performed in the point-form
of relativistic quantum mechanics, specifically focussing on the strange sector.
It is found that the relativistic 
predictions generally underestimate the experimental data. The nonrelativistic 
approximation of the approach leads to the decay operator of the 
elementary emission model. It is seen that the nonrelativistic reduction has
considerable effects on the decay widths.
\end{abstract}

\vspace{0.5cm}
There is already a long tradition in studying mesonic decays of baryon resonances
within constituent quark models (CQMs). However, most of the studies
hitherto have been performed within nonrelativistic or so-called relativised 
models~\cite{Stancu:1989iu,Capstick:1994kb,Geiger:1994kr,Ackleh:1996yt,
Theussl:2000sj}. Recently, the Graz group has presented relativistic CQM
calculations for $\pi$ and $\eta$ decays of $N$ and $\Delta$ resonances
employing a decay operator along the point-form spectator model (PFSM)
in the framework of relativistic (Poincar\'e-invariant) quantum
mechanics~\cite{Melde:2005hy}. A similar relativistic study
following a Bethe-Salpeter approach has been reported by the Bonn
group~\cite{Metsch:2003ix,Metsch:2004qk}. In this contribution we
report results for $\pi$ decays of strange 
baryon resonances by the relativistic Goldstone-boson exchange (GBE) 
and one-gluon exchange (OGE) CQMs of 
Refs.~\cite{Glozman:1998ag} and~\cite{Theussl:2000sj}, respectively.
The nonstrange decays of strange baryon resonances have not found much
attention in the past. So far there are no covariant results but only
the studies in Refs.~\cite{Capstick:1998uh,Krassnigg:1999ky,
Plessas:1999nb}. In addition to the relativistic predictions we also present
the decay widths resulting from the nonrelativistic reduction of the
PFSM decay operator, which corresponds to the elementary emission model (EEM).

\vspace{0.3cm}

The decay width of a baryon resonance is defined by the expression
\begin{equation}
\label{eq:decwidth}
{\mit\Gamma}_{i\to f}=\frac{|{\bf q}|}{4M^2}\;\frac{1}{2J+1}
\sum\limits_{M_J,M_{J'}}
\frac{1}{2T+1}\sum\limits_{\;M_T,M_{T'},M_{T_m}}
|F_{i\to f}|^2
\end{equation}
with the transition amplitude $F_{i\to f}$ given by the matrix element
of the reduced (four-momentum conserving) decay operator $\hat D^m_{rd}$
between incoming and outgoing baryon states
\begin{equation}
\label{eq:transel}
F_{i\to f}=
\langle V',M',J',M_{J'},T',M_{T'}|{\hat D}^m_{rd}
|V,M,J,M_J,T,M_T \rangle \, .
\end{equation}
Here, the index $m$ refers to the particular mesonic decay mode and
$q_\mu=(q_0,\bf q)$ denotes the four-momentum of the outgoing
meson in the rest-frame of the decaying baryon resonance
$|V,M,J,M_J,T,M_T \rangle$; the latter is characterized by the eigenvalues of the
velocity $V$, mass $M$, intrinsic spin $J$ with z-component $M_J$, and isospin
$T$ with z-projection $M_T$. Correspondingly the outgoing baryon state is
denoted by the primed eigenvalues. Representing the baryon eigenstates in a
suitable basis, the matrix element in Eq.~(\ref{eq:transel}) leads to the
integral
\begin{eqnarray}
&&
    {\langle V',M',J',M_{J'},T',M_{T'}|{\hat D}_{rd}^m|
    V,M,J,M_J,T,M_T\rangle}
    =
     {\frac{2}{MM'}\sum_{\sigma_i\sigma'_i}\sum_{\mu_i\mu'_i}{
	\int{
	d^3{\bf k}_2d^3{\bf k}_3d^3{\bf k}'_2d^3{\bf k}'_3
	}} } 
 \nonumber\\
&&
{\times \sqrt{\frac{\left(\sum_i \omega'_i\right)^3}
	{\prod_i 2\omega'_i}}
	\Psi^\star_{M'J'M_{J'}T'M_{T'}}\left({\bf k}'_i;
	\mu'_i\right)
	\prod_{\sigma'_i}{D_{\sigma'_i\mu'_i}^{\star \frac{1}{2}}
	\left\{R_W\left[k'_i;B\left(V'\right)\right]\right\}
	}
	} 
\nonumber\\
&&
     {\times
     \left<p'_1,p'_2,p'_3;\sigma'_1,\sigma'_2,\sigma'_3\right|{\hat D}_{rd}^m
	\left|p_1,p_2,p_3;\sigma_1,\sigma_2,\sigma_3\right>
	\nonumber } 
\nonumber\\
&&
	\times \prod_{\sigma_i}{D_{\sigma_i\mu_i}^{\frac{1}{2}}
	\left\{R_W\left[k_i;B\left(V\right)\right]\right\}}
	     {\sqrt{\frac{\left(\sum_i \omega_i\right)^3}
	{\prod_i 2\omega_i}}
	\Psi_{MJM_J TM_T}\left({\bf k}_i;\mu_i\right)} \, ,
\end{eqnarray}
where $\Psi_{MJM_J TM_T}\left({\bf k}_i;\mu_i\right)$ is the rest-frame 
wave function of the incoming baryon and analogously 
$\Psi^\star_{M'J'M_{J'}T'M_{T'}}\left({\bf k}'_i;\mu'_i\right)$  
the one of the outgoing baryon. Both wave functions result from the 
velocity-state representation of the baryon eigenstates. The momentum 
representation of the decay operator follows from the PFSM
construction~\cite{Melde:2005hy,Melde:2004qu}, 
where one assumes that only one of the quarks directly couples to the 
emitted meson:
\begin{multline}
\label{momrepresent}
\langle p'_1,p'_2,p'_3;\sigma'_1,\sigma'_2,\sigma'_3
|{\hat D}^{m}_{rd}|
 p_1,p_2,p_3;\sigma_1,\sigma_2,\sigma_3\rangle
 =-3{\left(\frac{M}{\sum_i{\omega_i}}
\frac{M'}{\sum_i{\omega'_i}}\right)^{\frac{3}{2}}
}\\
\frac{ig_{qqm}}{2m_1}\frac{1}{\sqrt{2\pi}}
\bar{u}(p_1',\sigma_1')\gamma_5\gamma^\mu \mathcal{F}^m 
 u(p_1,\sigma_1)q_\mu
 2p_{20}\delta^3\left(
{{\bf p}}_2-{{\bf p}}'_2\right)\delta_{\sigma_2 \sigma'_2}
2p_{30}\delta^3\left({{\bf p}}_3-{{\bf p}}'_3\right)
\delta_{\sigma_3 \sigma'_3} \, .
\end{multline}
Here, $g_{qqm}$ is the quark-meson coupling constant, $m_1$ the mass of the 
active quark, $\mathcal{F}^m$ the flavor-transition operator specifying the 
particular decay mode, and $u(p_1,\sigma_1)$ the quark spinor. All details 
of the formalism and the notation can be found in 
Ref.~\cite{Sengl:Diss}. The form of the decay operator is congruent with 
the calculations in Ref.~\cite{Melde:2005hy} and also consistent with the 
baryon charge normalisation and time-reversal invariance of the electromagnetic
form-factors~\cite{Wagenbrunn:2005wk}. The nonrelativistic approximation of 
the PFSM decay operator leads to the traditional EEM~\cite{Sengl:Diss}.
\renewcommand{\arraystretch}{1.4}
\begin{table}
{\footnotesize
\caption{Theoretical predictions for $\pi$ decay widths by the GBE and
OGE CQMs  in comparison to experiment~\cite{PDBook}. 
The relativistic calculations follow from the PFSM, 
while the EEM results represent their nonrelativistic limits. 
\label{tab1}
}
\begin{center}
{\begin{tabular}{@{}lcr| cccc| cccc@{}}
&&
&\multicolumn{4}{c|}{Theoretical Mass}
&\multicolumn{4}{c}{Experimental Mass}
\\
Decay &$J^P$&Exp. [MeV]
&\multicolumn{2}{c}{Relativistic}
&\multicolumn{2}{c|}{Nonrel. EEM}
&\multicolumn{2}{c}{Relativistic}
&\multicolumn{2}{c}{Nonrel. EEM}
\\
&&
& GBE & OGE & GBE & OGE  
& GBE & OGE & GBE & OGE 
\\
\hline
\multicolumn{1}{c}{\small $\rightarrow \Sigma\pi $}&
\multicolumn{5}{l}{}\\
\hline
$\Lambda(1405)$&
$\frac{1}{2}^-$&
$\left(50\pm2\right)$ 
    &$55$ 
    &$78$ 
    &$320$ 
    &$611$ 
    &$15$ 
    &$17$ 
    &$76$ 
    &$112$ 
\\ 
$\Lambda(1520)$&
$\frac{3}{2}^-$&
$\left(6.55\pm0.16\right)_{-0.04}^{+0.04}$ 
    &$5$ 
    &$9$ 
    &$5$ 
    &$8$ 
    &$2.8$ 
    &$3.1$ 
    &$2.1$ 
    &$2.3$ 
\\ 
$\Lambda(1600)$&
$\frac{1}{2}^+$&
$\left(53\pm38\right)_{-10}^{+60}$ 
    &$3$ 
    &$33$ 
    &$2$ 
    &$34$ 
    &$3$ 
    &$17$ 
    &$1.2$ 
    &$15$ 
%    &$14$
\\ 
$\Lambda(1670)$&
$\frac{1}{2}^-$&
$\left(14.0\pm5.3\right)_{-2.5}^{+8.3}$ 
    &$69$ 
    &$103$ 
    &$620$ 
    &$1272$ 
    &$68$ 
    &$94$ 
    &$572$ 
    &$1071$ 
\\ 
$\Lambda(1690)$&
$\frac{3}{2}^-$&
$\left(18\pm6\right)_{-2}^{+4}$ 
    &$19$ 
    &$25$ 
    &$24$ 
    &$28$ 
    &$18$ 
    &$21$ 
    &$23$ 
    &$22$ 
\\ 
$\Lambda(1800)$&
$\frac{1}{2}^-$&
$seen$ 
    &$68$ 
    &$101$ 
    &$473$ 
    &$1175$ 
    &$70$ 
    &$95$ 
    &$485$ 
    &$1095$
\\ 
$\Lambda(1810)$&
$\frac{1}{2}^+$&
$\left(38\pm23\right)_{-10}^{+40}$  
    &$3.8$ 
    &$2.1$ 
    &$55$ 
    &$150$ 
    &$4.1$ 
    &$5.0$ 
    &$55$ 
    &$94$ 
\\ 
$\Lambda(1830)$&
$\frac{5}{2}^-$&
$\left(52\pm19\right)_{-12}^{+11}$ 
    &$14$ 
    &$19$ 
    &$16$ 
    &$24$ 
    &$16$ 
    &$20$ 
    &$22$ 
    &$24$ 
\\ 
\hline 
$\Sigma(1385)$&
$\frac{3}{2}^+$&
$\left(4.2\pm0.5\right)_{-0.5}^{+0.7}$ 
    &$3.1$ 
    &$0.5$ 
    &$6.5$ 
    &$1.1$ 
    &$2.0$ 
    &$2.1$ 
    &$4.1$ 
    &$4.8$ 
\\ 
$\Sigma(1660)$&
$\frac{1}{2}^+$&
$seen$ 
    &$10$ 
    &$24$ 
    &$2$ 
    &$15$ 
    &$12$ 
    &$14$ 
    &$2.4$ 
    &$6.9$ 
\\ 
$\Sigma(1670)$&
$\frac{3}{2}^-$&
$\left(27\pm9\right)_{-6}^{+12}$ 
    &$15$ 
    &$23$ 
    &$21$ 
    &$32$ 
    &$13$ 
    &$17$ 
    &$17$ 
    &$21$
\\ 
$\Sigma(1750)^1$&
$\frac{1}{2}^-$&
$\left(3.6\pm3.6\right)_{-0}^{+5.6}$ 
    &$58$ 
    &$102$ 
    &$480$ 
    &$1249$ 
     &$63$ 
    &$102$ 
    &$574$ 
    &$1402$ 
    \\ 
$\Sigma(1750)^2$&
$\frac{1}{2}^-$&
$\left(3.6\pm3.6\right)_{-0}^{+5.6}$ 
    &$32$ 
    &$44$ 
    &$135$ 
    &$312$ 
    &$32$ 
    &$38$ 
    &$136$ 
    &$262$ 
\\ 
$\Sigma(1750)^3$&
$\frac{1}{2}^-$&
$\left(3.6\pm3.6\right)_{-0}^{+5.6}$ 
    &$10$ 
    &$1.0$ 
    &$116$ 
    &$34$ 
    &$10$ 
    &$0.9$ 
    &$110$ 
    &$32$ 
\\ 
$\Sigma(1775)$&
$\frac{5}{2}^-$&
$\left(4.2\pm1.8\right)_{-0.3}^{+0.8}$ 
    &$1.9$ 
    &$3.8$ 
    &$2.9$ 
    &$6.9$ 
    &$2.2$ 
    &$3.2$ 
    &$3.5$ 
    &$5.3$ 
\\ 
$\Sigma(1940)$&
$\frac{3}{2}^-$&
$seen$ 
    &$2.2$ 
    &$3.7$ 
    &$0.5$ 
    &$1.1$ 
    &$4.9$ 
    &$5.8$ 
    &$1.6$ 
    &$2.4$ 
\\ 
\hline
\multicolumn{1}{c}{\small $\rightarrow \Lambda\pi $}&
\multicolumn{5}{l}{}\\
\hline
$\Sigma(1385)$
&
$\frac{3}{2}^+$&
$\left(31.3\pm 0.5\right)_{-4.3}^{+4.4}$ &
$11$ &
$11$ & 
$25$ &
$28$ &
$14$ &
$13$ & 
$31$ &
$32$ 
\\ 
$\Sigma(1660)$
&
$\frac{1}{2}^+$&
$seen$ &
$8$ &
$5$&  
$6$ &
$0.02$ &
$10$ &
$3$&  
$8$ &
$0.05$
\\ 
$\Sigma(1670)$
&
$\frac{3}{2}^-$&
$\left(6\pm3\right)_{-1}^{+3}$ &
$2.5$ &
$2.0$& 
$5.5$ &
$5.1$&
$2.7$ &
$1.5$& 
$6.0$ &
$3.2$ 
\\ 
$\Sigma(1750)^1$
&
$\frac{1}{2}^-$&
$seen$ &
$1.6$ &
$1.5$& 
$43$ &
$67$ &
$0.8$ &
$1.4$& 
$49$ &
$70$ 
\\ 
$\Sigma(1750)^2$
&
$\frac{1}{2}^-$&
$seen$ &
$19$ &
$25$& 
$160$ &
$422$&
$18$ &
$25$ & 
$169$ &
$359$  
\\ 
$\Sigma(1750)^3$
&
$\frac{1}{2}^-$&
$seen$ &
$1.0$ &
$2.8$& 
$18$ &
$105$ &
$0.9$ &
$3$& 
$18$ &
$97$
\\ 
$\Sigma(1775)$
&
$\frac{5}{2}^-$&
$\left(20\pm 4\right)_{-2}^{+3}$ &
$6$ &
$10$& 
$10$ &
$21$ &
$8$ &
$8$& 
$15$ &
$15$ 
\\ 
$\Sigma(1940)$
&
$\frac{3}{2}^-$&
$seen$ &
$0.2$ &
$0.4$ & 
$1.7$ &
$3.5$ &
$0.5$ &
$0.5$ & 
$5.9$ &
$6.1$ 
\\ 
\hline
\multicolumn{1}{c}{\small $\rightarrow \Xi\pi $}&
\multicolumn{5}{l}{}\\
\hline
$\Xi(1530)$&
$\frac{3}{2}^+$&
$\left(9.9\right)_{-1.9}^{+1.7}$ &
$2.2$ &
$1.3$&
$4.4$ &
$3.0$&
$5.5$ &
$5.3$&
$11.4$ &
$12.5$ 
\\ 
$\Xi(1820)$
&
$\frac{3}{2}^-$&
$seen$ &
$0.4$ &
$1.6$&
$0.3$ &
$1.4$&
$0.7$ &
$1.2$&
$0.6$ &
$0.9$
\\
\hline
\end{tabular}}
\end{center}
}
\end{table}

In Table~\ref{tab1} we present the direct predictions of the GBE and OGE
CQMs for the $\pi$ decay modes of the strange baryon decay resonances
and compare with the latest compilation of the PDG~\cite{PDBook}.
Both the covariant PFSM results as well as the nonrelativistic EEM
results have been calculated with theoretical and experimental masses as
input. It is immediately evident that the relativistic predictions
usually underestimate the experimental data or at most reach them from below.
A similar finding was already made for the $\pi$ decay widths of $N$ and
$\Delta$ resonances~\cite{Melde:2005hy}. Here, there appear only two
exceptions, namely the widths of $\Lambda(1405)$ and $\Lambda(1670)$.
In case of the former it is caused by the (theoretical) mass, which
is far too high for both CQMs; the overprediction disappears when the
experimental mass is used. On the other hand the resonance mass of
the $\Lambda(1670)$ is more or less well reproduced in accordance with
experiment. In this case we may suspect the large decay width to
result from another reason, possibly a coupling of resonance states.

For the $\Sigma(1750)$ the CQMs offer three states that can be identified
with this resonance. In Table~\ref{tab1} we present the decay widths of
all theoretical levels (in the entries distinguished by the superscripts
1, 2, and 3). It is seen that the decay width of the third state
$\Sigma(1750)^3$ is pretty consistent with the magnitude of the
experimental data and it should be identified with the measured
$\Sigma(1750)$. The other two states can then be interpreted with
lower lying resonances (such as the $\Sigma(1620)$ and $\Sigma(1560)$)
not so well established by experiment. Regarding the classification of
these states see also Ref.~\cite{Melde:2006gg}.  

From the comparison of the PFSM results with experimental masses as input
one learns that the effects from different hyperfine interactions
generally play a minor role. Considerable influences are seen only in
$\Sigma\pi$ and $\Lambda\pi$ decays of $\Lambda(1600)$, $\Sigma(1750)^3$,
and $\Sigma(1660)$.

The nonrelativistic results corrresponding to the EEM scatter below and
above the experimental data. The effect of the nonrelativistic reduction
is strongly dependent on the decaying resonance. It is governed 
essentially by the truncation in the spin couplings as well as the
elimination of the Lorentz boosts.

We have reported the first covariant results for $\pi$ decays of strange
baryon resonances within CQMs. Obviously the approach needs further
improvements. In the first instance, one might think of a coupled-channel
formulation. The importance of additional Fock components has already
been seen in a PFSM calculation of mesons
decays~\cite{Krassnigg:2003gh,Krassnigg:2004sp} and also recent studies
of baryon resonances~\cite{Li:2005jn,Li:2006nm}. 

\vspace{0.3cm}

{\small This work was supported by the Austrian Science Fund (Projects 
FWF-P16945 and FWF-P19035).
B. Sengl acknowledges support by the Doktoratskolleg Graz
"Hadrons in Vacuum, Nuclei and Stars" (FWF-DK W1203).
The authors are grateful to L. Canton, A. Krassnigg, and 
R.~F. Wagenbrunn for useful discussions.}
{\footnotesize
\renewcommand\refname{}
\vspace{-1.4cm}

%\bibliographystyle{prsty}
%\bibliography{061013}

\begin{thebibliography}{10}

\bibitem{Stancu:1989iu}
F. Stancu and P. Stassart, Phys. Rev. D {\bf 39},  343  (1989).

\bibitem{Capstick:1994kb}
S. Capstick and W. Roberts, Phys. Rev. D {\bf 49},  4570  (1994).

\bibitem{Geiger:1994kr}
P. Geiger and E.~S. Swanson, Phys. Rev. D {\bf 50},  6855  (1994).

\bibitem{Ackleh:1996yt}
E.~S. Ackleh, T. Barnes, and E.~S. Swanson, Phys. Rev. D {\bf 54},  6811
  (1996).

\bibitem{Theussl:2000sj}
L. Theussl, R.~F. Wagenbrunn, B. Desplanques, and W. Plessas, Eur. Phys. J. A
  {\bf 12},  91  (2001).

\bibitem{Melde:2005hy}
T. Melde, W. Plessas, and R.~F. Wagenbrunn, Phys. Rev. C {\bf 72},  015207
  (2005); Erratum, ibid. to appear  .

\bibitem{Metsch:2003ix}
B. Metsch, U. Loering, D. Merten, and H. Petry, Eur. Phys. J. {\bf A18},  189
  (2003).

\bibitem{Metsch:2004qk}
B. Metsch, AIP Conf. Proc. {\bf 717},  646  (2004).

\bibitem{Glozman:1998ag}
L.~Y. Glozman, W. Plessas, K. Varga, and R.~F. Wagenbrunn, Phys. Rev. D {\bf
  58},  094030  (1998).

\bibitem{Capstick:1998uh}
S. Capstick and W. Roberts, Phys. Rev. D {\bf 58},  074011  (1998).

\bibitem{Krassnigg:1999ky}
A. Krassnigg {\it et~al.}, Few Body Syst. Suppl. {\bf 10},  391  (1999).

\bibitem{Plessas:1999nb}
W. Plessas {\it et~al.}, Few Body Syst. Suppl. {\bf 11},  29  (1999).

\bibitem{Melde:2004qu}
T. Melde, L. Canton, W. Plessas, and R.~F. Wagenbrunn, Eur. Phys. J. A {\bf
  25},  97  (2005).

\bibitem{Sengl:Diss}
B. Sengl, Dissertation, University of Graz, 2006.

\bibitem{Wagenbrunn:2005wk}
R.~F. Wagenbrunn, T. Melde, and W. Plessas, hep-ph/0509047  (2005).

\bibitem{PDBook}
W.-M. {Yao} {\it et~al.}, {J. Phys. G} {\bf 33},  1+  (2006).

\bibitem{Melde:2006gg}
T. Melde, W. Plessas, and B. Sengl, Bled Proceedings, Baryon Structure  (2006).

\bibitem{Krassnigg:2003gh}
A. Krassnigg, W. Schweiger, and W.~H. Klink, Phys. Rev. C {\bf 67},  064003
  (2003).

\bibitem{Krassnigg:2004sp}
A. Krassnigg, Phys. Rev. C {\bf 72},  028201  (2005).

\bibitem{Li:2005jn}
Q.~B. Li and D.~O. Riska, Phys. Rev. C {\bf 73},  035201  (2006).

\bibitem{Li:2006nm}
Q.~B. Li and D.~O. Riska, Phys. Rev. C {\bf 74},  015202  (2006).

\end{thebibliography}

}
\end{document}